\documentclass[12pt]{iopart}
\usepackage{amssymb}
\usepackage{graphicx}
\usepackage{color}      
\usepackage{multicol}
\usepackage{cite}
\usepackage{array}
\newcolumntype{L}{>{$}l<{$}}

\begin{document}

\title{Relaxation and Collective Excitations of Cluster Nano-Plasmas}

\author{Heidi Reinholz$^1$, Gerd R\"opke$^1$, Ingrid Broda$^2$, Igor Morozov$^{3,4}$, Roman Bystryi$^3$, Yaroslav Lavrinenko$^{3,5}$}

\address{$^1$ Institute of Physics, Rostock University, 18051 Rostock, Germany}  
\address{$^2$ Johannes Kepler Universit\"at Linz, Altenberger Strasse 69, 4040 Linz, Austria} 
\address{$^3$ Joint Institute for High Temperatures of RAS, Izhorskaya, 13, build. 2, 125412 Moscow, Russia}
\address{$^4$ National Research University Higher School of Economics, Myasnitskaya, 20, 101000 Moscow, Russia}
\address{$^5$ Moscow Institute of Physics and Technology (State University), Institutskiy per. 9, 141701 Dolgoprudny, Moscow Region, Russia}

\ead{heidi.reinholz@uni-rostock.de}

\begin{abstract}
Nano-plasmas produced, e.g.\ in clusters after short-pulse laser irradiation, can show collective excitations as derived from the time evolution of fluctuations in thermodynamic equilibrium.
Molecular dynamical simulations are performed for various cluster sizes. New data are obtained for the minimum value of the stationary cluster charge. 
The bi-local auto-correlation function gives the spatial structure of the eigenmodes, for which 
energy eigenvalues are obtained. Varying the  cluster size,
starting from a few-particles cluster, the emergence of macroscopic properties like collective excitations is shown.
\end{abstract}

\submitto{\jpb}
\maketitle

\section{Introduction}

The physics of nano-plasmas is presently a research field of increasing interest. 
Collective excitations of electrons in such plasmas are strongly coupled to external fields and influence  the energy deposition. 
With the increasing exploration of microstructures in recent technology developments, the understanding as well as the design of physical properties of nano-plasmas are of high interest. For example,  investigations of  near-field radiative heat transfer mediated by surface plasmon polaritons have  recently been performed, for a review see \cite{Boriskina-P15}. Femtosecond laser techniques  allow to study electron oscillations in ionized solid and gas clusters using the pump-probe method~\cite{Doeppner-PRA06,Fennel-PRL07,Hickstein-PRL14}.

As a special case, we consider metallic clusters such as sodium  of various sizes, 
characterized by the number $N$ of atoms. At solid state density, 
the atoms are singly ionized. For an electrically neutral cluster, 
the ion number $N_{\rm i} =N$ coincides with the number $N_{\rm e}$ of free electrons.
We assume that the ion density $n_{\rm i}$ in the clusters of different sizes remains constant
so that the volume $V_{\rm i}=n_{\rm i} N_{\rm i}$ of the respective cluster is proportional 
to the ion number $N_{\rm i}$. Then, in the thermodynamic limit $N_{\rm i}\to \infty$ the density $n_{\rm i}$ takes the bulk value.

In such clusters, a nonideal, strongly coupled nano-plasma is created by a laser pulse of 
 moderate intensity $I = 10^{12} - 10^{14}\,\mbox{W}/\mbox{cm}^2$, see \cite{Reinholz_PRB06}.
There is a large number of publications related to the experimental and theoretical 
investigation of the production and detection of excited clusters (see~\cite{ReinhardSuraud-book08,Fennel-RMP10}),  which we will not review here.  Our main interest is the  excitation of electrons in a quasi equilibrium nonideal nano-plasma generated in a metal cluster after irradiation by an intense femtosecond laser pulse, and the dynamical response
of the nano-plasma. 

The dynamical response of infinite, homogeneous nonideal plasmas is well 
known. The dynamical structure factor describes the correlation of density fluctuations in space and time. 
In particular, collective excitations such as plasmon resonances have been investigated 
theoretically within different approaches and are observed, 
e.g., within Thomson scattering experiments~\cite{Sperling-PRL15,Glenzer-RMP2009}.
According to the fluctuation-dissipation theorem, the absorption coefficient of radiation is 
related to auto-correlation functions in thermodynamic equilibrium. 
Analytical approaches, such as the Green function method, have been worked out 
to evaluate these quantities in the classical as well as in the quantum case, see, e.g.~\cite{Reinholz-ADP05}.
Alternatively, molecular dynamics (MD) simulations can describe classical plasmas at arbitrary 
coupling strength~\cite{Hansen-PRA81,Selchow-PRE01,NM-JETP05,RMRM-PRE04,MRRWZ-PRE05} 
where quantum properties of electrons are approximately accounted for by using 
specific electron-electron and electron-ion pseudopotentials, such as the 
Deutsch potential~\cite{Deutsch_PLA77}, the error-function (Erf) like 
potential~\cite{Zwicknagel-CPP03} or the so called corrected Kelbg 
potential~\cite{Filinov-JPA03-Kelbg}. 
Performing MD simulations, not only the equilibrium distributions in the phase space are obtained, 
but also fluctuations and their time dependence follow from the microscopic description.

For nano-plasmas, computer simulations are of even greater importance. 
Due to a relatively small number of particles the dynamical processes can be studied 
directly by atomistic simulations. Provided that the plasma temperature is high enough 
(few eV or more), the electron dynamics is governed mostly by the classical Coulomb forces. The method of classical MD simulations becomes one of the best choices in view of its simplicity and relatively high performance. It allows to account for particle collisions and goes beyond the simple equations for nano-plasma parameters~\cite{Ditmire_PRA96}. It is of importance for ionized metal clusters with the electron temperature of few eV where  nonideality effects are relevant.

Simulation methods with a more systematic treatment of quantum properties include wave packet 
molecular dynamics (WPMD)~\cite{Klakow-JCP94,GMMV-PRE13} or path-integral Monte Carlo 
(PIMC)~\cite{Filinov-PPCF01,Filinov-PRE15}. However, both methods have known problems that 
prevent their  direct application to simulations of bulk nonideal plasmas. For instance, the 
PIMC method suffers from the sign problem and high numerical efforts whereas the WPMD method 
has a problem with mapping quantum properties of electrons to classical quantities such as 
broadening of a free electron wave packet~\cite{MV-JPA09}. Here we present results of MD 
simulations where different electron-ion pseudopotentials have been used to implement quantum
effects.  The respective results are compared.

In the present paper we consider the dynamics of electrons after the formation 
of the nano-plasma, in particular the emission of electrons from the excited cluster 
(cluster charging), the relaxation of temperature, and the collective excitations of the
electron subsystem. 
The number of atoms $N$ is taken from 55 to $2 \cdot 10^4$ which allows to consider size effects. As an example  laser excited 
metal (sodium) clusters are considered. The properties to be discussed here are: i) the 
minimum charge of the cluster in a stationary state, ii) the electron temperature change due 
to the cluster ionization, and iii) the resonance frequencies of collective electronic excitations in the cluster.

The so called restricted molecular dynamics (RMD) has been proposed 
in Refs.~\cite{RRRM-IJMPB08,RRRMS-CPP09,RRRM-PRE11,BM-JPB15,Winkel-CPP13,Broda_CPP13,Raitza-NJP12} 
where the ion motion is neglected due to the large ion-electron mass ratio. 
When studying metal clusters the ionic subsystem can be treated either as a homogeneously 
charged full sphere~\cite{Broda_CPP13,Raitza-NJP12} which is called ``jellium model'' or as 
a crystal structure of point-like ions shaped by a 
sphere~\cite{RRRM-IJMPB08,RRRMS-CPP09,RRRM-PRE11,BM-JPB15,Winkel-CPP13}. 
In the second case, the ions are randomly displaced from their lattice sites to eliminate 
symmetry effects and averaging is performed over different ion distributions. 
Obviously, the RMD approach is a model assumption to work out the dynamics of the electron subsystem. 
In real situations, the ion configuration is changing and the excited cluster may expand,
so that the electron dynamics is coupled to the motion of the ions.

In Refs.~\cite{RRRM-IJMPB08,RRRMS-CPP09,RRRM-PRE11,BM-JPB15,Winkel-CPP13,Broda_CPP13,Raitza-NJP12}, the RMD method was used to study the properties of surface and volume plasmons in particular its dependence on the cluster size, plasma density, the ion ordering parameter and the specific type of the interaction potential. The dynamical properties of electrons and their response to external fields were described by correlation functions. Relaxation of the total cluster charge and temperature were considered. The standard MD simulation method has been used with direct all-to-all particle interactions.

Alternatively, the dynamical properties of the cluster nano-plasma can be computed using the MiCPIC method~\cite{Fennel-PRL12} which extends applicability of the classical MD to very large clusters (more than a million of atoms), or fluid dynamics \cite{Broda_CPP13,Gildenburg-PP2011}. With the help of contemporary computing facilities these methods allow to treat clusters with large numbers $N$ and to obtain results 
for the relaxation processes and collective excitations of nano-plasmas for a wider range of the cluster size.

This paper is organized as follows. In Sec.~\ref{sec:method} we review the computational method and plasma parameters. Then in Sec.~\ref{sec:charging} we discuss computational results and a simple theoretical model for the cluster charging. Finally in Sec.~\ref{sec:excitations} we consider electron excitations and the transition from cluster to bulk plasmas. The conclusions are drawn in Sec.~\ref{sec:conclusions}.

\section{Simulation method}
\label{sec:method}

For simplicity we consider a singly ionized hydrogen-like nano-plasma where the number of ions 
$N_\mathrm{i}$ is equal to the number of atoms $N$. The process of initial ionization of atoms 
or excitation of the electron subsystem by laser irradiation is beyond the scope of this work. According to the RMD scheme, instead of calculating the cluster expansion we consider the electron dynamics in the field of a fixed ion configuration.  The cluster expansion following a rapid ionization has been studied before in several papers, particularly in [26] using the same MD approach. Under the condition of moderate excitation (the temperature of few eV) considered here, the typical time of doubling the cluster size is hundreds of femtoseconds while the equilibration time for the electron subsystem is less than 10 fs. Therefore, the nano-plasma state can be assumed as near equilibrium and characterized by the electron temperature $T$ and the ion number density $n_\mathrm{i}$.  A more sophisticated approach should improve this adiabatic approximation taking the ion motion into account.

In other words, we consider a kind of snapshot of the cluster nano-plasma at a certain moment of expansion. As far as we put aside the ion dynamics we can calculate equilibrium MD trajectories for much longer than 100~fs and use time averaging instead of the ensemble averaging according to the ergodic theorem.

We consider both the jellium model as well as a configuration of point-like ions at position
$\{{\bf r}^{\rm ion}_i\}$. In the first case, a constant background ion density $n_\mathrm{i}$ is assumed without any fluctuations, while in the case of point-like ions, $n_\mathrm{i}$ denotes an average density. Along with $T$, $V$, $N_\mathrm{i}$, the number of electrons $N_\mathrm{e}$ is an additional, independent quantity to characterize the nano-plasma. The difference $Z = N_\mathrm{i} - N_\mathrm{e}$ determines the total cluster charge number.

There are different particle interaction models that account for quantum properties of electrons. In the case of point-like ions we follow Refs.~\cite{Zwicknagel-CPP03,RRRM-IJMPB08,RRRMS-CPP09,RRRM-PRE11,BM-JPB15,Calisti-HEDP11} where the electron-ion interaction potential is given by
\begin{equation}\label{eq:poterf}
U_\mathrm{Erf}(r) = - \frac{e^2}{4\pi\epsilon_0\,r}
  \mathrm{Erf}\left( \frac{r}{\lambda} \right),
\end{equation}
with a parameter $\lambda$ described below. This model corresponds to the Gaussian shape of an electron wave function which is used also in WPMD simulations.

Within the jellium model, the ions are replaced by the effective potential~\cite{Broda_CPP13}
\begin{equation}\label{eq:potjellium}
U_\mathrm{Jellium}(r)=
\left\{ \begin{array}{@{\kern2.5pt}lL}
\frac{e^2 n_{\mathrm{i}}}{\epsilon_0} \left(\frac{r^2}{6}-\frac{R_\mathrm{i}^2}{2}\right), &$\mathrm{for}~r<R_\mathrm{i}$, \\
- \frac{e^2\,N_\mathrm{i}}{4\pi\epsilon_0\,r}, &$\mathrm{else~where}$,
\end{array}\right.
\end{equation}
where $R_\mathrm{i} = \left(3N_\mathrm{i}/(4\pi\,n_\mathrm{i})\right)^{1/3}$ is the cluster radius. Here the pure Coulomb interaction between electrons and ions is assumed.

For the electron-electron interaction we consider different models such as the pure Coulomb repulsion
\begin{equation}\label{eq:potcoul}
U_\mathrm{Coul}(r) = \frac{e^2}{4\pi\epsilon_0\,r}.
\end{equation}
and the repulsive Erf-like potential $-U_\mathrm{Erf}(r)$ from~(\ref{eq:poterf}). The value of the parameter $\lambda_\mathrm{ei}$ in~(\ref{eq:poterf}) for the electron-ion interaction is taken to be $0.318$~nm which provides the correct ionization energy for a sodium atom, $I_\mathrm{P} = U_\mathrm{ei}^\mathrm{erf} (r \rightarrow 0) = e^2 / (2 \pi^{3/2} \lambda_{\mathrm{ei}} \epsilon_0) = -5.1$~eV. Following~\cite{RRRM-IJMPB08,RRRMS-CPP09,RRRM-PRE11,BM-JPB15}, for the electron-electron interaction the same value of $\lambda_\mathrm{ee} = \lambda_\mathrm{ei}$ is used.

In some works a potential similar to~(\ref{eq:poterf}) is used where the $\lambda$ parameter denotes the de Broglie wavelength $\Lambda$ (see, e.g.~\cite{Zwicknagel-CPP03}). For the plasma parameters used here ($T=2.2$~eV and $n_\mathrm{e} = 2.7 \times 10^{22}\:\mathrm{cm}^{-3}$) we obtain $\Lambda = 0.62$~nm while the average distance between the electrons is $r_\mathrm{ee} \approx 0.24$~nm. Thus the plasma degeneracy parameter $\Theta=k_\mathrm{B}T/\epsilon_{\mathrm{F}} = 2 m_\mathrm{e} k_{\mathrm{B}} T_\mathrm{e} (3 \pi^2 n_\mathrm{e})^{-2/3} \hbar^{-2} = 0.67$ is less than unity. It formally conflicts with the classical MD approach but our bulk plasma simulations show that the MD simulations with the given pseudopotential are still valid up to the values of $\Theta \approx 0.4$ (see~\cite{Lavrinenko-JPCS17}).

For the initial positions of ions $\{{\bf r}^{\rm ion}_i\}$, we use either a random distribution within the sphere $R_\mathrm{i}$ or a regular mesh. In any case, the initial interparticle distance is assured to be not less than $r_\mathrm{min} = (3/(2\pi n_\mathrm{i}))^{1/3}$. The electrons are initially  placed on top of ions or distributed randomly. The procedure of equilibration of the electron gas is considered in the next section.

The leap-frog integration scheme has been used for solving the equations of motion for electrons  with a typical time step of $10^{-18}$~s or less. Using this time step ensures the total energy conservation with an accuracy of 1\% for $5 \cdot 10^7$ MD steps. The total MD trajectory length ranges from 9.0~ps to 0.5~ns depending on the cluster size. The simulation results are averaged over $4-6$ statistically independent MD trajectories with different initial positions of ions (in the case of a random ion distribution). The GPU accelerated code is used to speed up simulations and increase the averaging accuracy.

\section{Charge and Temperature Relaxation}
\label{sec:charging}
\indent
The RMD simulations allow to study the dynamics of the electrons under 
``external'' conditions specified by the cluster size $R_\mathrm{i}$, 
the ion density $n_\mathrm{i}$ and the temperature $T$ of the electron subsystem.
The specific electron temperature can be obtained by using the Langevin thermostat during the first stage 
of the simulations. If the motion of the electrons is not restricted, sufficient energy to escape from the cluster  can be allocated to the electrons (emitting process). After the emission of an electron, 
the cluster increases its total charge $Ze$ by the value of $e$ 
which results in an increased cluster Coulomb potential. 

The increase of the cluster potential reduces the probability of further electron emission. 
From the theoretical point of view, the emission of electrons stops when the kinetic energy of 
all the electrons becomes smaller than the cluster potential. In practice, the 
equilibration process can be stopped when the cluster charge is stable for a sufficiently long 
time (nanosecond and more). When this state is reached the thermostat can be switched off. 
Then, the electron dynamics is analyzed for quasi-equilibrium conditions. Using this method 
in Refs.~\cite{RRRM-IJMPB08,RRRMS-CPP09,RRRM-PRE11,BM-JPB15} 
the time evolution of the electron subsystem has 
been computed in dependence on the above mentioned parameter values $R_\mathrm{i}$, 
$n_\mathrm{i}$, and~$T$.

This approach, however, is not applicable for studying the cluster charging itself because the 
thermostat affects the relaxation process by depositing some extra energy  
which compensates the energy loss due to electron emission. As the energy deposition is hard 
to control, the final cluster charge and the relaxation time turn out to be dependent on the 
thermostatting process.

The cluster charging itself can be studied using an alternative technique proposed in 
Refs.~\cite{Broda_CPP13,Raitza-NJP12}. Here, the electrons are constrained within the sphere 
$R_\mathrm{i}$ by a special wall potential during the equilibration. In this case the 
electrons rather quickly gain an equilibrium state with a given temperature $T_0$ while all of 
them stay inside the cluster, $N_{\rm e}(t<t_0)=N_{\rm i}$. 
At $t_0$, the thermostat is switched off, and the constraining wall is removed. The 
subsequent relaxation of the electron subsystem is analyzed.

In the present section, 
we discuss the simulation of the resulting relaxation processes for $T>t_0$ owing to this latter approach. Because electrons escape when the confining wall is removed, the cluster charge number $Z$ increases. As shown in Ref.~\cite{Broda_CPP13}, the emission rate is slowing down,
and if the charge state of the cluster $Z(t)$ remains constant for a time interval larger than 
1~ns, we consider this as a stationary (quasi equilibrium) state with 
a charge number $Z_{\rm fin}$. 
Further emission of electrons is possible but is considered as rare process 
compared with the time scale of the electron subsystem dynamics governed by 
the plasma frequency $\omega^{}_\mathrm{pl} = \sqrt{e^2 n_\mathrm{e}/(\epsilon_0 m_\mathrm{e})}$.
For the sodium bulk value $n_\mathrm{e}=n_\mathrm{i} = 2.7 \times 10^{22}\:\mathrm{cm}^{-3}$ 
we have $\omega^{}_\mathrm{pl}=9.26\:\mathrm{fs}^{-1}$.

\begin{figure}[ht]
		\centering
		\includegraphics[width=7.0cm]{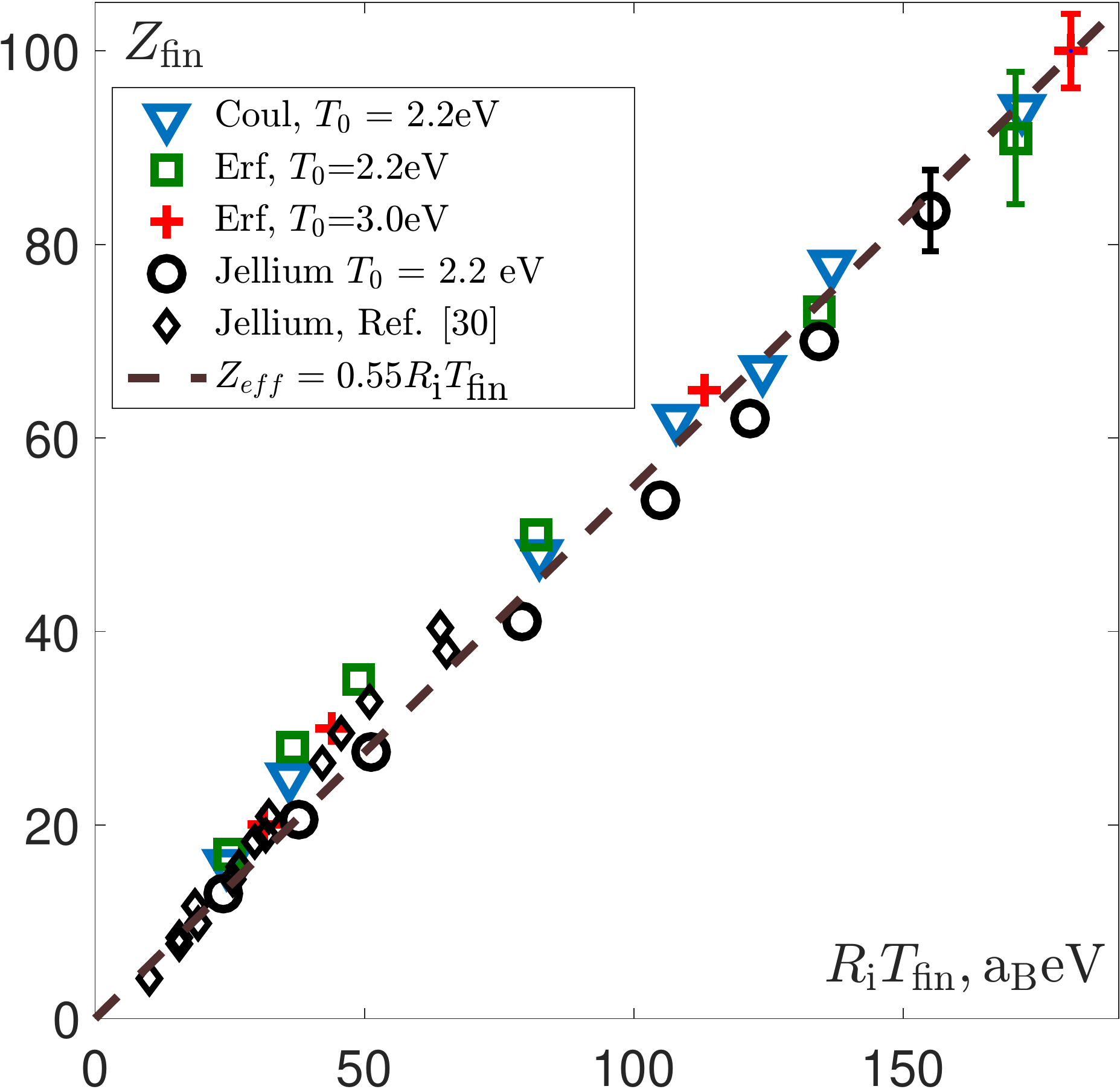}
		\caption{\label{Fig:Charge}The final cluster charge $Z_\mathrm{fin}$ in dependence on the product of the cluster radius $R_\mathrm{i}$ and the final electron temperature $T_\mathrm{fin}$. Symbols represent MD results with different interaction potentials and initial temperatures $T_0$. Previous results ($\diamondsuit$) are taken from Ref.~\protect\cite{Broda_CPP13}. Dashed line represents the linear fit according to Eq.~(\protect\ref{eq:final_charge}).}
\end{figure}

\begin{figure}[ht]
		\centering
		\includegraphics[width=7.1cm]{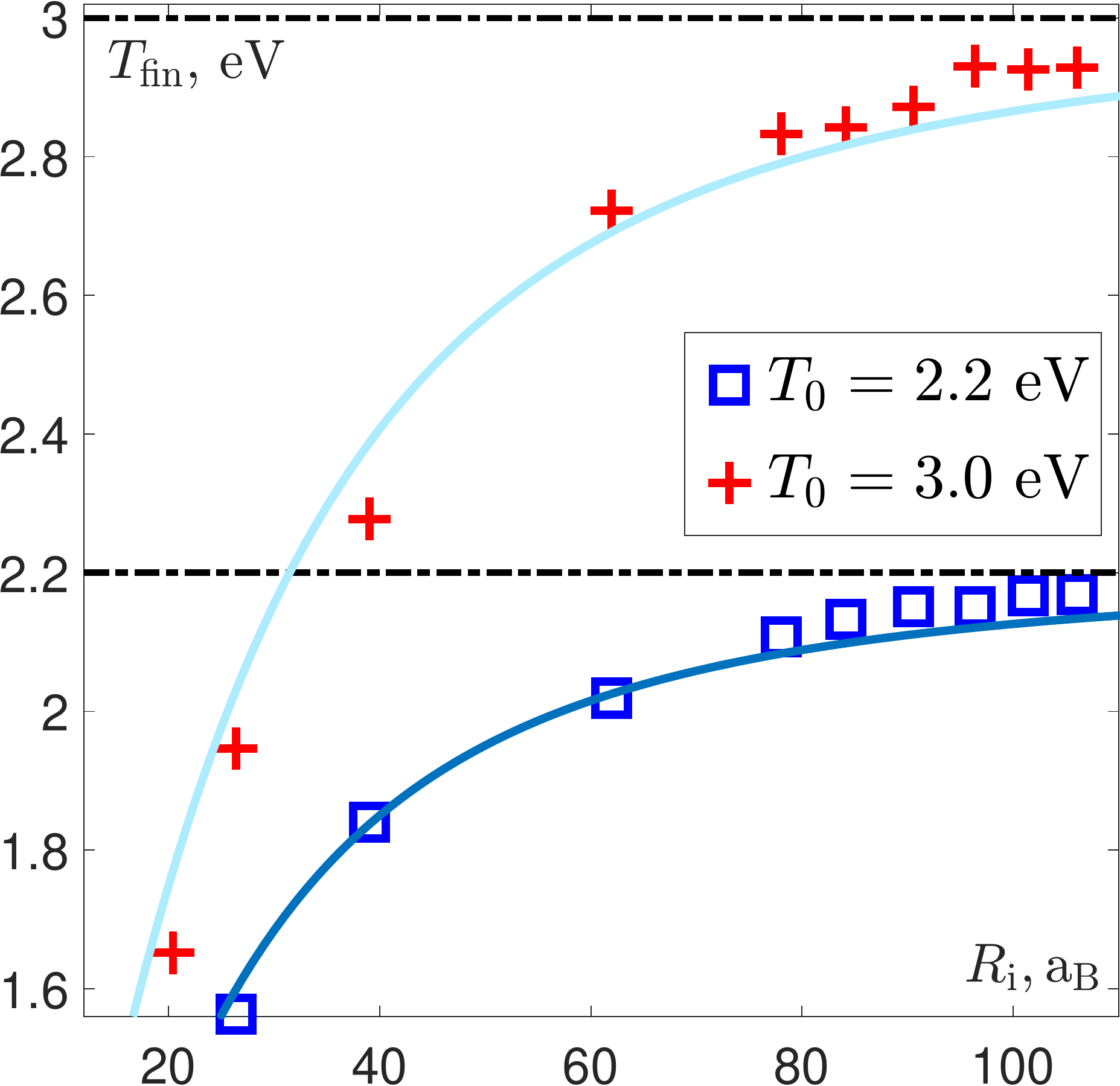}
		\caption{\label{Fig:FinalTemp}Final temperature $T_\mathrm{fin}$ of electrons confined in the cluster  as function of the cluster size $R_\mathrm{i}$. Symbols represent MD results with the error function potential (\ref{eq:poterf}) and different initial temperatures $T_0$. Curves represent the solution of Eq.~(\protect\ref{eq:final_temperature}).}
\end{figure}

We consider cluster nano-plasmas with the ion density $n_\mathrm{i} = 2.7 \times 10^{22} \mathrm{cm}^{-3}$. Figs.~\ref{Fig:Charge} and~\ref{Fig:FinalTemp} present the final charge $Z_\mathrm{fin}$ and the final temperature $T_\mathrm{fin}$ obtained in a quasi equilibrium state for different cluster models (jellium and point-like ions) and different types of electron-electron interaction potentials. Two values of the initial temperature $T_0 = 2.2$~eV and $T_0 = 3.0$~eV are used which correspond to the values of coupling parameter $\Gamma = e^2/(4 \pi \epsilon_0 k_BT) (4 \pi n_\mathrm{e}/3)^{1/3}$ of $3.17$ and $2.32$, respectively (at the beginning the electron number density is $n_\mathrm{e} = n_\mathrm{i}$).

As seen from Fig.~\ref{Fig:Charge} the final charge of the cluster $Z_\mathrm{fin}$ is strongly dependent on the initial temperature $T_0$ and the cluster size $R_\mathrm{i}$. However, all the results can be fitted by a simple relation (see~\cite{Broda_CPP13})
\begin{equation}\label{eq:final_charge}
Z_\mathrm{fin} = c_1 R_\mathrm{i}\, k_{\rm B} T_\mathrm{fin}
\end{equation}
with the parameter value of $c_1=(0.55 \pm 0.06)\:(a_\mathrm{B}\,\mathrm{eV})^{-1}$, where $a_{\rm B}$ is the Bohr radius. This value is in good agreement with the result of~\cite{BM-JPB15}, while it is slightly lower than $c'_1=0.65\:(a_\mathrm{B}\,\mathrm{eV})^{-1}$ as found  in Ref.~\cite{Broda_CPP13}.

In the bulk limit $N_\mathrm{i} \to \infty$, the cluster charge is increasing, 
$Z_{\rm fin}\propto N_\mathrm{i}^{1/3}$. As a surface effect, we have, however,
$\lim_{N_\mathrm{i} \to \infty} Z_{\rm fin}/N_\mathrm{i} = 0$ 
for any fixed $n_\mathrm{i}$ and $T_\mathrm{fin}$. 

As demonstrated in Fig.~\ref{Fig:Charge}, the different models with respect to the interaction
or the ion configuration have no significant influence on the relation (\ref{eq:final_charge})
which underlines that it is determined mainly by the average field of the screened cluster
 potential in which the electrons are moving. The short-range behavior of the interaction
is also not of significance. The minimum charging $Z_\mathrm{fin}$ can take large values  
and would be of interest for nano-structure plasmas.
A more detailed discussion of the general form of
relation (\ref{eq:final_charge}) is found in Ref.~\cite{Broda_CPP13}.

Another interesting signature of the cluster charging after the deconfinement at $t_0$ is the 
change of temperature because the emitted electrons extract kinetic energy from the 
excited cluster so that the final temperature of the cluster is lower than the 
initial temperature $T_0=T(t_0)$, i.e. $T_\mathrm{fin}<T_0$.
The dependence of the final temperature on the cluster size, see in Fig.~\ref{Fig:FinalTemp}, shows that the relative energy loss decreases with  increasing cluster radius $R_\mathrm{i}$. It can be described analytically using the following considerations.

We assume that the number of evaporated electrons $Z$ is small compared to the number of electrons in the nano-plasma $N_\mathrm{e} $ but it is greater than one, i.e. $1 \ll Z \ll N_\mathrm{e} $. It means that the potential barrier caused by the cluster charge $Ze$ that prevents the remaining electrons from escape is given by the potential at the cluster surface,
\begin{equation}\label{eq:Clust-pot}
  U_\mathrm{b} = \frac{Z e^2}{4\pi\epsilon_0\,R_\mathrm{i}}.
\end{equation}
The value of this potential should be larger than the electron thermal energy, e.g.\  $U_\mathrm{b} \gg k^{}_\mathrm{B} T$, to consider the evaporation as a slow process. A further condition follows from Eq.~(\ref{eq:final_charge}) and reads as $c_1 R_\mathrm{i}  k_{\rm B} T << N_\mathrm{e}$ which restricts applicability of this model to rather low temperatures.

Within these assumptions we can estimate the average energy taken away by a single emitted electron as $U_\mathrm{b}$. Therefore the change of the total kinetic energy of electrons in the nano-plasma can be written as
\begin{equation}\label{eq:Cons-law}
\frac{d}{dt}\left(\frac{3}{2} N_\mathrm{e} k^{}_\mathrm{B} T\right) = - \frac{dZ}{dt} U_\mathrm{b},
\end{equation}
where $T$ is an instantaneous temperature. The previous simulation results~\cite{RRRM-IJMPB08,RRRMS-CPP09} show that the typical time between electron emissions is much greater than the time for the local equilibrium build up in the electron subsystem.

Assuming that the change of $N_\mathrm{e}$ is small with respect to the change of $Z$, the temperature $T_\mathrm{fin}$ at a time $t_\mathrm{fin}$ can be obtained via integration
\begin{equation}\label{eq:main_after_integration}
k^{}_\mathrm{B} (T_\mathrm{fin} - T_0)
= k^{}_\mathrm{B}\int\limits_{T_0}^{T_\mathrm{fin}}\!d T
= -\frac{2}{3N_\mathrm{e}} \int\limits_{0}^{Z_\mathrm{fin}}\!
  \frac{Z e^2}{4\pi\epsilon_0\,R_\mathrm{i}} dZ
= -\frac{Z_\mathrm{fin}^2 e^2}{12\pi\epsilon_0\,R_\mathrm{i}N_\mathrm{e}},
\end{equation}
where $Z_\mathrm{fin} = Z(t_\mathrm{fin})$ is the final number of emitted electrons. If we use  the empirical relation~(\ref{eq:final_charge}) between $Z_\mathrm{fin}$ and $T_\mathrm{fin}$, and substitute $N_\mathrm{e}=\frac{4}{3}\pi R_{\mathrm{i}}^3 n_\mathrm{e}$, then the final temperature can be expressed as
\begin{equation}\label{eq:final_temperature}
T_\mathrm{fin}
= T_0 - \frac{e^2 c_1^2}{16 \pi^2 \epsilon_0  n_\mathrm{e}}\,
  \frac{k^{}_\mathrm{B} T_\mathrm{fin}^2}{R_\mathrm{i}^2}
= T_0 - c_2 k^{}_\mathrm{B}\left( \frac{T_\mathrm{fin}}{R_\mathrm{i}} \right)^2,
\end{equation}
where the parameter $c_2 = e^2 c_1^2 /(16 \pi^2 \epsilon_0  n_\mathrm{e})$ 
is introduced for convenience. For the nano-plasma considered above it is equal to 
$c_2 = (165 \pm 35)\; a_{\rm B}^2/\mathrm{eV}$. For large clusters, i.e.\ large values
of $R_\mathrm{i}$, the temperature remains nearly unchanged during the emission process of hot electrons. As seen from Fig.~\ref{Fig:FinalTemp}, 
the theoretical curves obtained from the solution of the quadratic 
equation~(\ref{eq:final_temperature}) are in a good agreement with the MD results.

\section{Collective excitations}
\label{sec:excitations}

\subsection{Total current autocorrelation function}

The properties of nano-plasmas are quite different compared to the homogeneous case
owing to the long range character of the Coulomb interaction.
Surface effects become important. However, if the finite system becomes sufficiently large,
the bulk properties should dominate, and the asymptotic limit of the homogeneous case is expected for $N_{\rm i} \to \infty$.

Similar to the homogeneous plasma case~\cite{Selchow-PRE01,RMRM-PRE04,MRRWZ-PRE05}, we study collective excitations of electrons by analyzing the total current-current correlation function (current ACF). With the local current micro-density
\begin{equation}
{\bf j}({\bf r},t)=\sum_{c={\rm e,i}}\sum_{\alpha}^{N_c} e_c{\bf v}_{c,\alpha}(t)
\,\delta({\bf r}_{c,\alpha}(t)-{\bf r})
\end{equation}
we obtain the $\bf k$-dependent current micro-density ($\Omega_0$ denotes the normalization volume)
\begin{equation}
{\bf j}_{\bf k}(t)=\frac{1}{\Omega_0}\int d^3r \,{\bf j}({\bf r},t)\,
e^{i {\bf k} \cdot {\bf r}}.
\end{equation}
In particular, ${\bf j}_{\bf k=0}(t)$ is the total current density. A related quantity is the charge micro-density
\begin{equation}
{\varrho}({\bf r},t)=\sum_{c={\rm e,i}}\sum_{\alpha}^{N_c} e_c
\delta({\bf r}_{c,\alpha}(t)-{\bf r}),
\end{equation}
where also a  $\bf k$-dependent  charge micro-density is introduced by Fourier transformation.
The macroscopic density is obtained by averaging the micro-densities over the equilibrium ensemble as denoted by $\langle \dots \rangle$.

The current density (in $\bf r$ or $\bf k$ space) is a vector field which can be decomposed in a longitudinal ($L$) and a transverse ($T$) component. The longitudinal component is related to the charge density via the equation of continuity, 
\begin{equation}
 \frac{\partial}{\partial t} \langle {\varrho}({\bf r},t) \rangle 
+ {\rm div}\langle {\bf j}({\bf r},t) \rangle =0.
\end{equation}

We introduce the normalized auto-correlation function (ACF) of the total current density
\begin{equation}
\label{ACFjj}
K_{j_0j_0}(t)^{L/T}=\frac{1}{\langle j_0^2 \rangle} \frac{1}{t_\mathrm{avr}}\int_0^{t_\mathrm{avr}}
({\bf j}_{0}(t_0+t)^{z/x} {\bf j}_{0}(t_0))^{z/x}\, dt_0,
\end{equation}
where the ensemble average is replaced by the time average as noted above, and $t_{\rm avr}$ is the averaging time. In our MD simulations we calculate ACF for the interval of $0 \le t \le t_\mathrm{max}$, $t_\mathrm{max} \approx 1$~ps, and average it over $5\cdot 10^6$ initial points $t_0$ for $N_\mathrm{i}=55$ (the total trajectory length is $t_\mathrm{avr} = t_\mathrm{traj} = 5$~ns) and $1.4\cdot 10^3$ initial points for $N_\mathrm{i}=55$ ($t_\mathrm{traj} = 9$~ps, 4 trajectories are use for averaging).

Note that for $\bf k=0$, the longitudinal and transverse components $L/T$ are different because in performing the limit $\bf k \to 0$, surface charges have to be taken into account (see the calculations in \cite{MRRWZ-PRE05}). Plain surfaces are assumed perpendicular to the $z$ direction, and the $z$ component of the current density gives the longitudinal part.

The dimensionless Fourier transform of the current ACF
\begin{equation}\label{current-acf}
K^{L/T}(\omega) = \frac{\omega^{}_\mathrm{pl}}{4\pi} \int K^{L/T}(t) \mathrm{e}^{i\omega t} dt,
\end{equation}
can be analyzed in terms of a generalized Drude formula which defines the dynamical collision frequency according to the relation
\begin{equation}\label{genDrude}
\frac{\nu(\omega)}{\omega_{\rm pl}}=\frac{1}{K^L(\omega)}
+i\left( \frac{\omega}{\omega_{\rm pl}}-\frac{\omega_{\rm pl}}{\omega}\right),
\end{equation}
see Refs.~\cite{RMRM-PRE04,MRRWZ-PRE05}. It should be underlined that we do not employ the standard Drude formula with a constant relaxation time. Only the structure of Eq. (\ref{genDrude}) is motivated by the Drude formula,
but the dynamical collision frequency defined here has the full information of the response function and, in this way, also of the dielectric function. The frequency dependence implies that this is a complex function where the imaginary part is connected with the real part by a Kramers-Kronig relation.

For the bulk nonideal plasma, good agreement between analytical calculations using a Green-function approach to the dynamical collision frequency $\nu(\omega)$ and MD simulations has been shown for small up to moderate plasma coupling strengths~\cite{MRRWZ-PRE05}. It should be mentioned that, in contrast to the analytical quantum statistical calculations which are  appropriate only for $\Gamma \leq 2$, the MD simulations are applicable for an arbitrary coupling strength, but consider quantum effects only within the pseudopotential approach. 

Coming back to finite clusters, the auto-correlation function (\ref{ACFjj}) 
has been evaluated for different model systems.
As the particle interaction potential is a crucial input parameter for MD simulations we considered different interaction models as listed in  Table~\ref{tab:potentials}, see Eqs.~(\ref{eq:poterf})--(\ref{eq:potcoul}) for definitions.

\begin{table}[h]
\begin{center}
\caption{\label{tab:potentials}Interaction models between different sorts of particles used in MD simulations of cluster nano-plasmas.} 
\begin{tabular}{|c|c|c|c|}
	\hline
	label & ion-ion & electron-ion & electron-electron\\
	\hline
	Jellium & --- & $U_\mathrm{Jellium}$ & $U_\mathrm{Coul}$ \\
	Erf & $U_\mathrm{Coul}$ & $U_\mathrm{Erf}$ & $-U_\mathrm{Erf}$ \\
	Coul & $U_\mathrm{Coul}$ & $U_\mathrm{Erf}$ & $U_\mathrm{Coul}$ \\ 
	\hline
\end{tabular}
\end{center}
\end{table}

\begin{figure}[htp]
	\centering
	\includegraphics[width=0.45\linewidth]{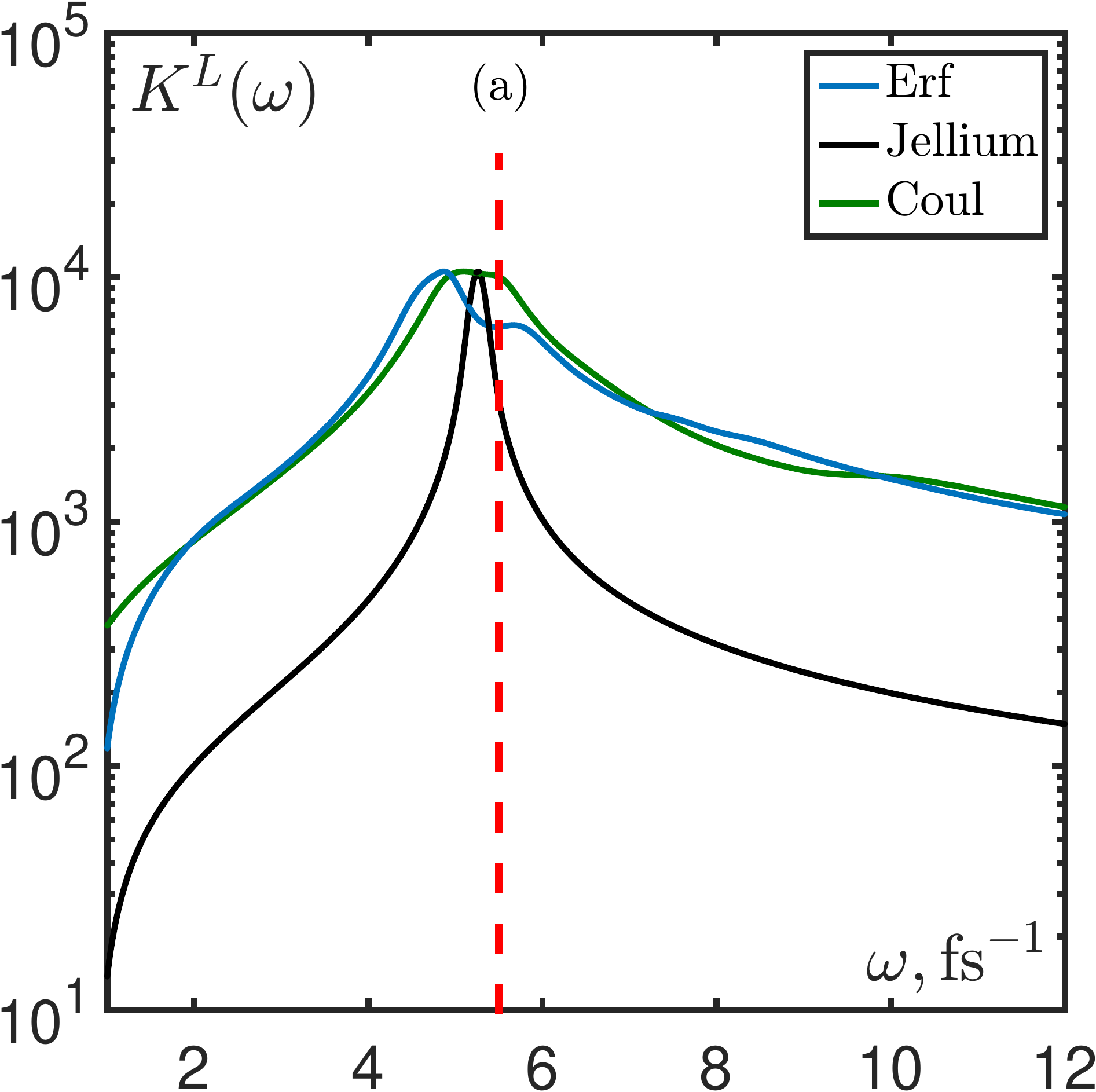}\quad
	\includegraphics[width=0.4\linewidth]{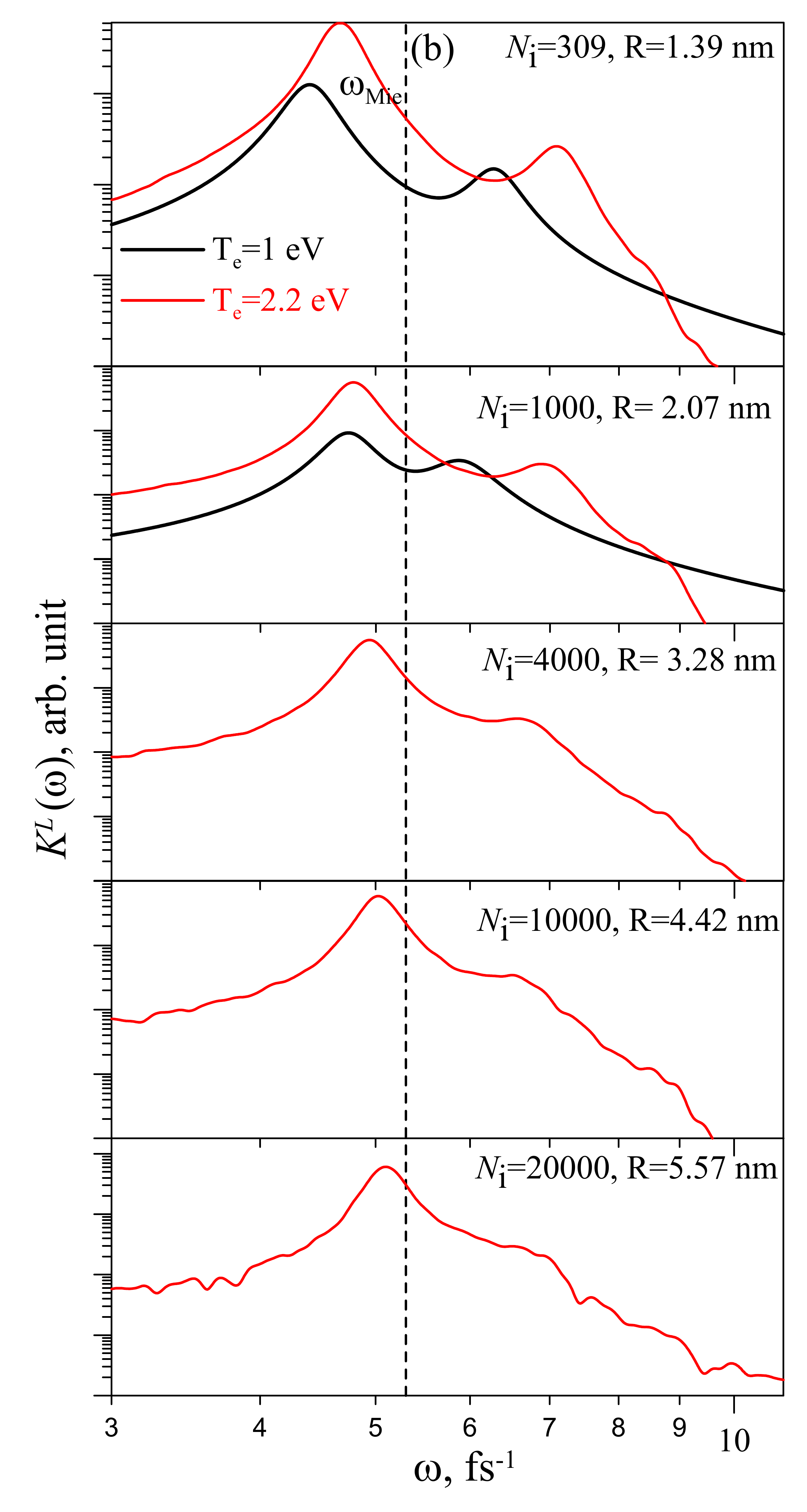}
	\caption{\label{fig:acf}
Fourier transform of the total current ACF~(\protect\ref{current-acf}) for (a) a cluster with $N_\rmi=55$ and electron temperature $T_\mathrm{e} = 2.2$~eV using different particle interaction models, and (b) different cluster sizes and  electron temperatures (x-axis: logarithmic frequency scale). The electron density is equal to $n_e = 2.7 \times 10^{22}~\rm{cm}^{-3}$ for all cases. The  Mie frequency, shown by the dashed lines, is equal to $\omega^{}_{\mathrm{Mie}}=5.35\:\mathrm{fs}^{-1}$. }
\end{figure}

Examples of the Fourier transform of the current ACF obtained from MD simulations with different interaction models are shown in Fig.~\ref{fig:acf}a. The results are obtained using the RMD approach similar to Refs.~\cite{RRRM-IJMPB08,RRRMS-CPP09,RRRM-PRE11,BM-JPB15}. Several peak structures are seen, while a main peak is near the Mie frequency $\omega_\mathrm{Mie}^2=\omega_\mathrm{pl}^2/3$, as expected. It is seen that for the ``Jellium'' model the damping of the Mie oscillations is small as electron-ion collisions are missing. For both ``Erf'' and ``Coul'' models the peak is broadened and red-shifted with respect to the Mie frequency.

The redshift is studied extensively in Refs.~\cite{RRRM-PRE11,BM-JPB15,Winkel-CPP13}. The ACF spectra obtained in Ref.~\cite{Winkel-CPP13} contain additional peaks which could be due to  peculiarities in the averaging process or during  numerical Fourier transformation used in this work. The results of~\cite{RRRM-PRE11} and \cite{BM-JPB15} coincide for small clusters while in~\cite{BM-JPB15} the simulations have been performed for a wider range of ion numbers ($N_i=55 - 10^5$) extending the results towards the bulk plasma limit.

In the course of preparing this paper, we revisited the results of~\cite{BM-JPB15} where it was reported that for small clusters only the Mie (surface) mode is seen in the current ACF while for large clusters the plasmon-like (volume) mode appears and becomes increasingly predominant with  cluster growth. We found that the appearance of the plasmon-like mode resulted from the use of a fixed-size sphere for the ACF calculation, so that for $N_\mathrm{i}>2000$ the cluster size is larger than this sphere. A similar effect was reported in Ref.~\cite{LMV-JPCS16},  where plasmon-like oscillations were obtained for a subcell inside a periodical simulation cell. In the present work, we considered the complete cluster volume when calculating the ACF (see Fig.~\ref{fig:acf}b).  We found that the Mie peak does not change its amplitude significantly  with  cluster growth and no plasmon-like peak is seen in the ACF at $\omega^{}_\mathrm{pl}=9.26\:\mathrm{fs}^{-1}$.

\begin{figure}[h]
	\begin{center}
		\includegraphics[width=8cm]{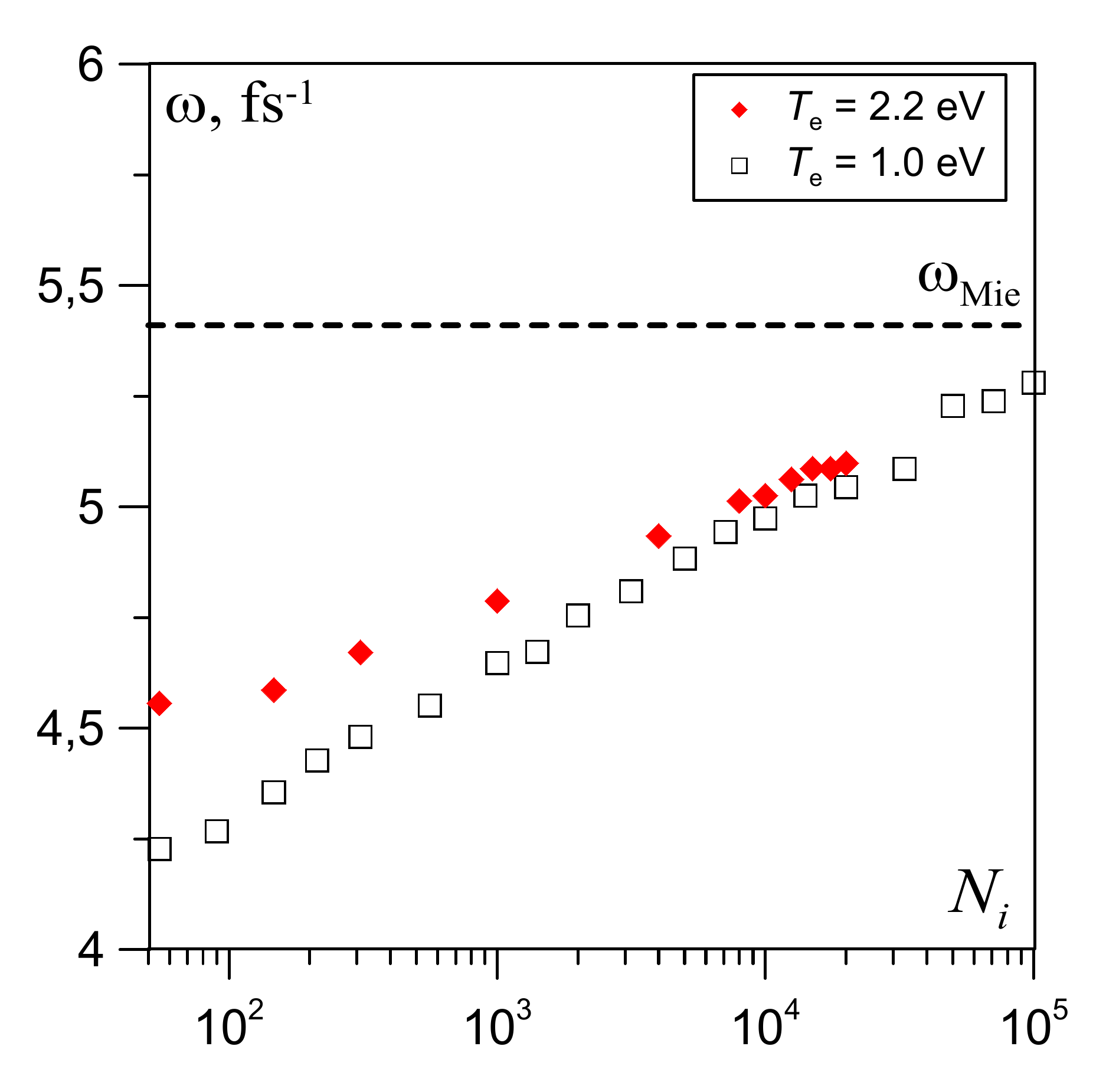}
	\end{center}
	\caption{\label{fig:freq-size}
The Mie resonance frequencies depending on the number of ions in the cluster, for different electron temperatures: $\diamondsuit$ -  $T_\mathrm{e} = 2.2$~eV (this work), $\Box$ -   $T_\mathrm{e} = 1.0$~eV Ref.~\cite{BM-JPB15}.}
\end{figure}

The dependence of the Mie resonance frequency on the cluster size is shown in Fig.~\ref{fig:freq-size}. Our new results are close to those reported in~\cite{BM-JPB15}. As discussed in Refs.~\cite{RRRM-PRE11,BM-JPB15}, the redshift of the Mie resonance is a consequence of the electron density distribution within the cluster. 
Instead of a hard-sphere distribution for which the Mie formula applies, the radial distribution of the electron 
density declines to zero continuously and is washed out at the cluster surface. With increasing 
cluster size, this surface effect becomes less relevant, and the value of the Mie resonance 
approaches the value for the one of a macroscopic homogeneous sphere.

However, further structures are seen in the total current ACF, shown in Fig.~\ref{fig:acf}, which can 
hardly be resolved.  Therefore, a more detailed investigation of the electron dynamics 
in the excited cluster is necessary to analyze these signals.

\subsection{Electron excitations in cluster nano-plasmas}

For  bulk matter, a consistent description of collective excitations is  given satisfactorily using a plane wave with the wave vector ${\bf k}$. In contrast to this homogeneous case,  excitations in finite systems have a more complex structure determined by the surface boundary conditions.
The Fourier transform of the total current ACF allows to reveal resonances in nano-plasmas but does not show the nature of different excitations. For this the spacial structure of the excitations should be analyzed.

We extract the current density ${\bf j}({\bf r},t)$ from the MD simulation run and calculate the Laplace transform
$\hat K_{jj}({\bf r},{\bf r}',\omega)$
of the time-averaged current density auto-correlation function
\begin{equation}
\hat K_{jj}({\bf r},{\bf r}',\tau)=\lim_{T_\mathrm{avr} \to \infty} \frac{1}{T_\mathrm{avr}} \int_0^{T_\mathrm{avr}}  {\bf j}({\bf r},t){\bf j}({\bf r}',t+\tau)\, dt,
\end{equation}
which coincides with the ensemble average according to the ergodic principle. This bi-local current ACF
$\hat K_{jj}$ is a tensor, but we consider only the $j_z-j_z$ component. It 
is symmetric and can be diagonalized by solving the eigenvalue problem
\begin{equation}\label{bi-local-acf}
\sum_{a'=1}^{N_{\rm cells}} {\rm Re}\,K_{jj}({\bf r}_a,{\bf r}_{a'},\omega) \Psi_\mu({\bf r}_{a'},\omega)=
K_\mu(\omega) \Psi_\mu({\bf r}_{a},\omega)
\end{equation}
at fixed frequency $\omega$. In simulations, we decompose the space into $N_{\rm cells}$ cells which are denoted by $a$. In each cell the average current density is calculated. Spherical coordinates are used. 

\begin{figure}[h]
	\begin{center}
	 \includegraphics[width=12cm]{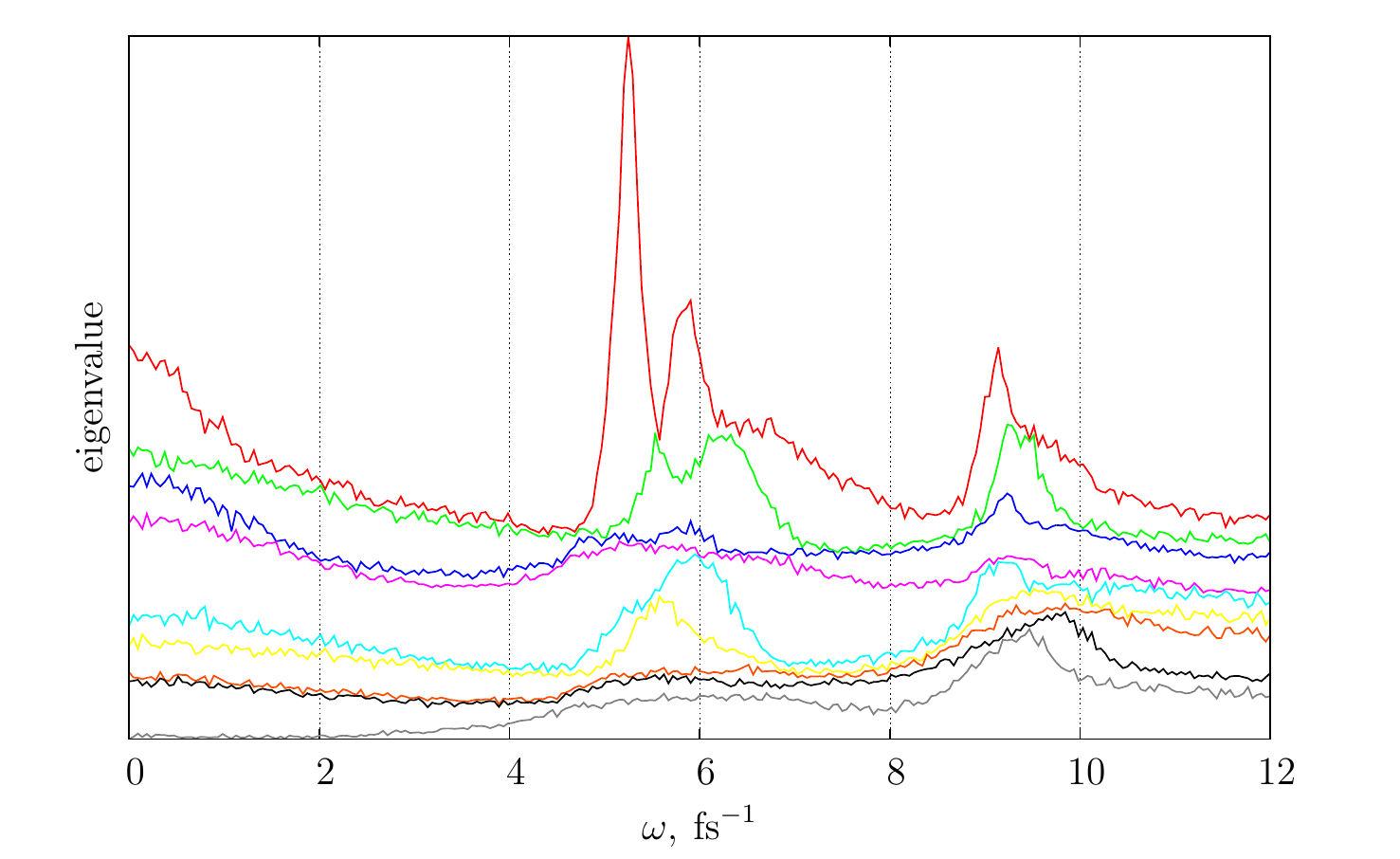}
	\end{center}
	\caption{\label{fig:eigenvalues}Eight strongest eigenvalues of the bi-local current ACF~(\protect\ref{bi-local-acf}), as a function of frequency, using the jellium model for a  Na$_{309}$ cluster at $T=2.2$~eV.}
\end{figure}

\begin{figure}[h]
	\begin{center}
		\includegraphics[width=12cm]{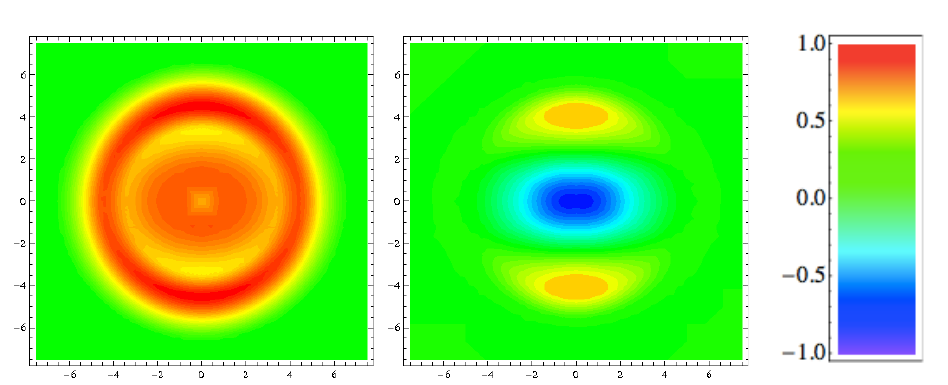}
	\end{center}
	\caption{\label{fig:eigenmodes}Eigen modes of electronic excitations in nano-plasmas, see also \cite{RRRM-PRE11}, for a
	 Na$_{309}$ cluster at $T=1.96$ eV, $n=5.2 \times 10^{21}$ cm$^{-3}$. Amplitude (color code, in arbitrary units) of the current ACF's spacial structure  in the $z-x$ plane (in units of $a_{\rm B}$) at a fixed azimuthal angle (on which it does not depend) for the main collective surface mode (left) and the plasma-like mode (right). }
\end{figure}

The eigenvalues $K_\mu(\omega)$ as function of the frequency $\omega$ are shown in 
Fig.~\ref{fig:eigenvalues}. Sharp 
resonances are obtained. The eigenvalues are ordered by the largest absolute value at a frequency under consideration. Their frequency dependence leads to crossings where a different mode becomes leading, see \cite{RRRM-PRE11} for details.
Modes which do not have  a large dipole moment do become visible in the mode spectrum allthough  they can  not be clearly seen
in Fig.~\ref{fig:acf} . 
The corresponding eigenmodes $\Psi_\mu({\bf r}_{a},\omega)$ are given by the eigen functions. 
An example for the mode analysis is shown in Fig.~\ref{fig:eigenmodes}. Here it is seen that we have surface modes such as the Mie mode, where  the current density is nearly constant in the entire volume, but at the surface the charge density is changing. Furthermore, we have volume plasmons which 
are expected to be the precursor of the plasmon modes in the homogeneous plasma.

The widths of the peaks seen in Fig.~\ref{fig:eigenvalues} correspond to the damping of the corresponding excitation modes. Preliminary results for the damping rates for the surface and volume plasmons were published in Ref.~\cite{BM-JPB15}. In order to distinguish contributions from the collisional and Landau damping (see the results for thin films in Ref.~\cite{Zaretsky_JPB04}) additional processing is required which has not been done yet. An example of such processing for a bulk plasma can be found in Ref.~\cite{NM-JETP05}.

\section{Conclusions}
\label{sec:conclusions}

New results of MD simulations for clusters of different size have been presented to show the transition from small clusters to homogeneous systems. To obtain stationary conditions for clusters at finite temperature, the charging has to be considered. Results for the minimum charging are well described by a simple formula~(\ref{eq:final_charge}). It would be of interest whether linear relations similar to expression~(\ref{eq:final_charge}) are also valid for non-spherical nano-plasmas.

Another interesting result is the cooling owing to the charging process because hot electrons are emitted. The difference $T_0-T_\mathrm{fin}$ is decreasing with increasing cluster size. A relation for $T_\mathrm{fin}(T_0,R_\mathrm{i})$ has been found which fits well with simulation data. 

The collective excitations of nano-plasmas are of high interest because the coupling to external fields is very efficient. We discussed the bi-local correlation functions and identified the eigenmodes, the excitation energy as well as their spacial structure. The deviation of the dipole excitation frequency from the Mie excitation energy  is due to the inhomogeneous electron density distribution inside the cluster, in contrast to the Mie model where a constant density is assumed for the electrons inside the cluster.

Previously published results for the excitation spectra depending on the cluster size~\cite{BM-JPB15} are corrected while the dependence of the Mie-like oscillation mode frequency on the cluster size is confirmed. A detailed discussion of the collective excitations and the corresponding spacial structure of the eigenmodes has been given in \cite{RRRM-PRE11}. We  note that similar modes occur also in nuclei where, beside the giant dipole resonance, also further excitation modes are known. Further properties such as the Thomas-Reiche-Kuhn sum rule are of interest. 

Calculation of the current bi-local ACF for a quasi equilibrium cluster nano-plasma (see~\cite{RRRM-IJMPB08,RRRMS-CPP09,RRRM-PRE11,BM-JPB15,Winkel-CPP13,Broda_CPP13}) provides information about electron excitations that correspond to the absorption resonances of irradiated clusters. 
 Different bi-local correlation functions can be investigated such as the density-density correlation function which is related to the longitudinal current bi-local ACF, see \cite{RRRM-PRE11,Broda-CPP12}. Rotations and vorticies are obtained from the transversal part of the current density ACF.  We found that in contrast to the jellium model, the motion on the fluctuation field of point-like ions makes the Mie-like resonance much broader.
The damping of the collective excitations is an interesting aspect for future work.

\ack
The authors are thankful to Thomas Fennel for fruitful discussions. I.~Broda, H.~Reinholz and G.~R\"opke acknowledge the Collaborative Research Center (DFG) 652. I.~Morozov, R.~Bystryi and Y.~Lavrinenko acknowledge the support from the Russian Foundation for Basic Research, grant No.~15-02-08493a. Computations were performed on the cluster K-100 (KIAM RAS, Moscow) and the Supercomputer Centre JIHT RAS.



\section*{References}

\bibliography{Reinholz_jphysb}

\end{document}